\documentclass[11pt,a4paper]{article}
\usepackage{latexsym}
\usepackage{graphicx}
\usepackage[cp1251]{inputenc}
\usepackage{subfigure}
\usepackage{amsmath}
\usepackage{amsfonts,amssymb}
\usepackage{indentfirst}

\begin{document}

\title{\bf QUATERNION REPRESENTATION AND SYMPLECTIC SPIN TOMOGRAPHY}
\maketitle
\begin{center}

\author{Aleksey K. Fedorov$^{1,2\,*}$ and Evgeny O. Kiktenko$^{2,3}$}
\vspace{0.25cm}

{\small 
$^{1}${\it Russian Quantum Center, Novaya St. 100, Skolkovo, Moscow Region, 143025, Russia. \\}
$^{2}${\it Bauman Moscow State Technical University, 2nd Baumanskaya St., 5, Moscow 105005, Russia. \\}
$^{3}${\it Geoelectromagnetic Research Center of Schmidt Institute of Physics of the Earth, Russian Academy of Sciences, P.O. 30 Troitsk, Moscow Region 142190, Russia}}
\end{center}

{Corresponding author e-mail}: akf@rqc.ru; {e-mail}: evgeniy.kiktenko@gmail.com

\vspace{0.5cm}
Quantum tomography for continuous variables is based on the symplectic transformation group acting in the phase space. 
A particular case of symplectic tomography is optical tomography related to the action of a special orthogonal group. 
In the tomographic description of spin states, the connection between special unitary and special orthogonal groups is used. 
We analyze the representation for spin tomography using the Cayley--Klein parameters and discuss an analog of symplectic tomography for discrete variables. 
We propose a representation for tomograms of discrete variables through quaternions and employ the qubit-state tomogram to illustrate the method elaborated.
\vspace{0.5cm}


\section{Introduction}

Using quantum systems as a basis for information theory and information technologies provides new interesting possibilities. 
As is well known, in classical computing the basic unit is a bit with two values \cite{Shannon}. 
In quantum computation, the superposition state called qubit is used, and several problems can be efficiently solved using the quantum algorithms introduced by Shor \cite{Shor}. 
Problems related to complete characterization of quantum states and quantum processes are very important.

For example, the problem of appropriate description of quantum states both in theory and experiment is very important. 
The description of quantum states purely in probability terms looks like a natural generalization of the Shannon theory in a quantum domain. 
Recently, much attention has been paid to quantum tomography \cite{Manko}-\cite{Lvovsky3}. 
In quantum tomography, the states are described in terms of normalized nonnegative probability-distribution functions. 
In addition, in \cite{Manko}-\cite{Manko10} it was shown that quantum tomograms are directly related to well-known quasiprobability distributions \cite{Wigner}-\cite{Glauber2}.

On the other hand, the search for the best experimental realization of qubits is an important problem. 
It is remarkable that tomograms are measurable. 
In superconducting circuits, quantum tomograms were used for characterization of quantum states of current (voltage) in the Josephson junction \cite{Manko4}.
In quantum optics, tomograms are measured by homodyne detection \cite{Beck}-\cite{Lvovsky}.
A review of the latest developments in quantum-state tomography of optical fields, including applications, was presented in \cite{Lvovsky}.. 
A more complicated issue is quantum process tomography \cite{Lvovsky2}-\cite{Lvovsky3}.

As a result, thanks to quantum tomography, we have a unified view on quantum information. 
The tomographic approach to quantum systems with continuous variables was developed in \cite{Manko}-\cite{Manko2}, 
while different problems in this representation were considered [\ref{Manko7}-\ref{Manko9}, \ref{Fedorov}, \ref{Fedorov2}, \ref{Filippov}].
The tomographic description of systems with discrete variables was proposed in [\ref{Manko3}-\ref{Manko6}, \ref{Manko9}, \ref{Manko10}, \ref{Manko11}, \ref{DAriano}, \ref{Filippov}].
In our work, we concentrate on the tomographic representation of quantum systems with discrete variables. 

More precisely, we consider the tomographic representation for spin states, being interested in spin tomograms related to the irreducible representation of the ${\rm SU(2)}$ group, as a particular case of unitary tomograms related, respectively, to the irreducible representation of the ${\rm SU(N)}$ group. 
The spin tomogram, in fact, is the probability distribution function depending on parameters of the $U\in{\rm SU(2)}$ group \cite{Manko2}-\cite{Manko5}. 
In view of a natural relationship between the rotation group ${\rm SO(3, \mathbb{R})}$ in the three-dimensional Euclidean space $\mathbb{R}^{3}$ and ${\rm SU(2)}$, these parameters are given by Euler angles.

In this paper, we present a rather simple and conceptual modification of the spin-tomography scheme based on the other parameterization of the ${\rm SU(2)}$ group, namely, we suggest parameterization through unit quaternions, which form a symplectic group ${\rm Sp(1)}$ related to ${\rm SO(3, \mathbb{R})}$ and ${\rm SU(2)}$ groups. 
Hypercomplex numbers as the basis for the description of spin states was introduced in [\ref{Hamilton}].. 
Apparently, this approach is less physically implementable but more potentially useful in numerical calculations in the case of large spin systems, e.g., in the Ising-like model. 
The quaternion representation of spin tomogram is an analog of symplectic tomography in the domain of discrete variables.

This paper is organized as follows.
In Sec. 2, we discuss a general scheme and some group-theoretical aspects of quantum tomography, 
where we review the case of continuous variables mostly based on theoretical results of \cite{Manko}-\cite{Manko2} 
and experimental schemes of \cite{Beck}-\cite{Lvovsky} and give a brief introduction to spin tomography with the theoretical conception of \cite{Manko2}-\cite{Manko5} and experimental implementation of \cite{DAriano}). 
In Sec. 3, we present spin tomograms in terms of quaternion parameterization and use the qubit state as an example. We conclude with a brief summarization of our results.

\section{Group Theory Aspects of the Quantum Tomogrpahy}\label{General}

Quantum states in a Hilbert space $\mathcal{H}$ are associated with positive Hermitian operators $\hat{\rho}$ with unit trace $\mathrm{Tr}\hat{\rho}=1$.
Let $\Omega(\mathcal{H})$ be a set of quantum states in $\mathcal{H}$.
Quantum tomograms are given by a mapping of the element $\hat\rho\in\Omega(\mathcal{H})$ on a parametric set of the probability distribution functions realized by the transformation
\begin{equation}\label{map}
	\hat\rho\in{\Omega(\mathcal{H})}\xrightarrow{\text{G}(g)}\mathcal{T}\{g,m\},
\end{equation}
where $m$ is a physical observable, ${\rm G}(g)$ is a transformation group with parametrization by $g$, and parametric set $\mathcal{T}\{g,m\}$ is called quantum tomogram of the state $\hat\rho_m$.

Elements of the tomogram are marginal distribution functions.
When $m$ is continuous variable we get a scheme for continuous variables tomography \cite{Manko}-\cite{Manko2}. 
There are two cases of parametrization: in the first case $g$ is discrete, and in the second case $g$ is continuous and given in functional form.
If $m$ is discrete we get a scheme for discrete variables tomography \cite{Manko2}-\cite{Manko5}. 
And we also have two cases: parametrization $g$ is discrete (optimal schemes) or parametrization $g$ is continuous \cite{Filippov}.

There are several important circumstances.
First, the transformation given by ${\rm G}$ is the group transformation. 
Usually, ${\rm G}$ is a classical group. 
This is remarkable, because in this case the parameterization of the group is well known and, in addition, the group structure of transformation is important due to its known properties (this will be considered in detail further). Second, mapping (\ref{map}) defines the quantization procedure and the reverse procedure in a natural way. 
These problems were considered for continuous and discrete variables in [\ref{Manko8}, \ref{Filippov}]. 
Third, we can consider the action of ${\rm G}$ as the action of the state as well as the action of an observable. 
Finally, there exists a good physical interpretation --- every element of the parametric set $\mathcal{T}\{g,m\}$ is a probability of observing the value m after transformation ${\rm G}$.

\subsection{Tomography of Continuos Variables}\label{Continuos}

In this section we consider tomographic mapping (\ref{map}) for continuous variables in more detail \cite{Manko}.
The key point of quantum description for continuous variables is commutation relation between canonical position $\hat{q}$ and momentum $\hat{p}$ operators.
Mathematically, commutation relation is a skew-symmetric bilinear form, which analogical to Poisson bracket is classical mechanics.
Thus, group $G$ is transformation group, preserving these forms, i.e. transformation of variables is canonical.

\subsubsection{Symplectic Tomography}

Canonical transformations of the symplectic manifold on the phase space $\Phi$ are given by the $\rm{Sp}(2n,\mathbb{R})$ group of symplectic transformations of the phase space with $\dim{\Phi}=2n$. 
The group $\rm{Sp}(2n,\mathbb{R})$ is the set of $2n\times2n$ matrices with standard matrix multiplication. 
Transformation from one set of canonical variables $(q,p,t)$ to the other one $(Q,P,t)$ is a canonical iff the Jacobi matrix of this transformation is the symplectic one. 
There exists a useful representation of the $\rm{Sp}(2,\mathbb{R})$ group in the matrix form
\begin{equation}\label{transformation}
	\begin{pmatrix}
	\hat{Q} \\
	\hat{P} \\
	\end{pmatrix}=
	\begin{pmatrix}
	\mu & \eta \\
	\acute{\eta} & \acute{\mu} \\
	\end{pmatrix}\\
	\begin{pmatrix}
	\hat{q} \\
	\hat{p} \\
	\end{pmatrix},
\end{equation}
where the set $\{\mu, \eta, \acute{\mu}, \acute{\eta}\}\in\mathbb{R}$ is a parameterization of the $\rm{Sp}(2,\mathbb{R})$ group matrix (\ref{transformation}), and the determinant of the matrix is unity. 
Thus, mapping (\ref{map}) of the density matrix to the family of probability-distribution functions is given by the relation
\begin{equation}\label{def}
	\mathcal{T}(Q, \mu, \eta)={\rm Tr}\{\hat\rho\delta(Q-\mu\hat{q}-\eta\hat{p})\}=\langle{Q, \mu, \eta}|{\hat\rho}|{Q, \mu, \eta}\rangle, \quad \hat\rho\in\Omega(\mathcal{H}),
\end{equation}
where $|{Q, \mu, \eta}\rangle$ is an eigenvector of the Hermitian operator $\mu\hat{q}+\eta\hat{p}$ for the eigenvalue $Q$.

In integral form, relation (\ref{def}) reads
\begin{equation}\label{density}
	\mathcal{T}(Q, \mu, \eta)=\frac{1}{(2\pi)^3}\int{\hat\rho\left(q+\frac{\xi}{2},q-\frac{\xi}{2}\right)\exp(-ip\xi)\delta(Q-\mu{q}-\eta{p})dqdpd\xi},
\end{equation}
where $\delta$ is the Dirac delta-function. 
This representation is closely connected with the Weyl and star-product quantization \cite{Manko9}. 
Recall that the Weyl symbol of the density matrix is explicitly the Wigner function connected with symplectic tomogram. 
The Weyl representation determined by the projective representation of the commutative translation group and relation (\ref{density}) is connected with this fact.

\subsubsection{Optical Tomography}

The procedure of balanced homodyne photon detection is based on mixing of a measurable (weak) field and a strong coherent field with varying phase $\theta$ on the beam splitter. 
In this case, the measurable observable is $\hat{Q}=\widehat{q}\cos\theta+\widehat{p}\sin\theta$. 
The angle $\theta$ could be interpreted as a rotation angle of the phase space. 
Therefore, our consideration of canonical transformation (\ref{transformation}) is reduced to considering the matrix of the ${\rm SO}(2, \mathbb{R})$ group
\begin{equation}\label{optical}
	\begin{pmatrix}
	\cos{\theta} & \sin{\theta}  \\
	-\sin{\theta} & \cos{\theta}  \\
	\end{pmatrix}\\,
\end{equation}
This group is parameterized by  $\theta\in\mathbb{R}/2\pi\mathbb{Z}$.

Relations (\ref{transformation})-(\ref{density}) for symplectic tomogram are transformed to equivalent relations for optical tomograms if $\mu=\cos{\theta}$, $\eta=\sin{\theta}$, $\acute{\eta}=-\sin\theta$, and $\acute{\mu}=\cos\theta$.
Symplectic tomograms provide a most general case of the probability-distribution function in quantum tomography. 
Note that the symplectic tomogram is a function of two parameters of the $\rm{Sp}(2,\mathbb{R})$ group parameterization, and the optical tomogram is a function of the parameter $\theta$. 
Therefore, we can conclude that the symplectic tomography formalism is useful in theory but sophisticated for implementation in practice.

\subsection{Tomography of Spin States}\label{Spin}

A complete diagram of discrete-variables tomography schemes in connection with different quantizaion types was suggested in \cite{Filippov}. 
In the case of a system with discrete variables, mapping (\ref{map}) and the definition of the tomogram (\ref{def}) are transformed to the following relation:
\begin{equation}\label{defd}
	\mathcal{T}_m(U)=\langle{m}|U\hat\rho{U}^{\dagger}|{m}\rangle, \quad \hat\rho\in\Omega(\mathcal{H}),
\end{equation}
and the positivity and normalization of the tomogram follow directly, namely,
\begin{equation}\label{Normalization}
	\sum\nolimits_m{\mathcal{T}_m(U)}=1, \qquad {\mathcal{T}_m(U)}\geq 0.
\end{equation}

If the matrix $U\in{\rm SU(2)}$, then (\ref{defd}) is a general definition for spin tomogram. 
In the case of $U\in{\rm SU(N)}$, this probability representation of spin states is called unitary tomography \cite{Filippov}.

\subsubsection{Spin Tomogram and the Cayley--Klein Parameters}

Any matrix $U\in{\rm SU(2)}$ in (\ref{defd}) has the following form:
\begin{equation}
	U=\begin{pmatrix}
	\alpha & \beta \\
	-\beta^{*} & \alpha^{*} \\
	\end{pmatrix}, 
\end{equation}
where $\alpha,\beta\in\mathbb{C}$ are the Cayley--Klein parameters, and the following relation holds $|\alpha|^2+|\beta|^2=1$. 
These numbers provide the ${\rm SU(2)}$-group parameterization. 
We can rewrite a pair of the Cayley--Klein parameters in the form $\alpha=\alpha_1+i\alpha_2$ and $\beta=\beta_1+i\beta_2$. As an result, we obtain
\begin{equation}\label{defd2}
	\mathcal{T}_m(\alpha_1, \alpha_2, \beta_1, \beta_2)=\langle{m}|U(\alpha_1, \alpha_2, \beta_1, \beta_2)
	\hat\rho{U^{\dagger}(\alpha_1, \alpha_2, \beta_1, \beta_2)}|{m}\rangle, \quad \hat\rho\in\Omega(\mathcal{H}),
\end{equation}

Further detailed consideration of the spin tomography scheme is based on the method of $U\in{\rm SU(2)}$ parameterization and using its irreducible representations. 
Here we can see that the tomogram is defined as a function of four real parameters: $\alpha_1$, $\alpha_2$, $\beta_1$, and $\beta_2$.

\subsection{Spin Tomogram as Function of the Euler angles}

Starting from pioneer work in the field \cite{Manko2}, parameterization through the Euler angles was suggested as a physically natural alternative to (\ref{defd2}). 
The use of the Euler angles is one of the ways to represent the rotation of $\mathbb{R}^3$. 
A more general statement is that the rotations of $\mathbb{R}^3$ form a group isomorphic to ${\rm SO(3)}$. 
To prove this statement, consider a rotation transformation in $\mathbb{R}^3$. 
It is clear that rotation in the space is the transformation $\mathbf{x}'=\Lambda\mathbf{x}$ of the element ${\bf x}=(x_1,x_2,x_3)\in\mathbb{R}^3$ 
with preservation of the Euclidean metric $||\mathbf{x}||$ and zero point (we use bold type for vectors), 
where $\Lambda$ is an orthogonal matrix, which forms a group$\rm O(3)$ under matrix multiplication. 
As a result of the obvious relation for the orthogonal matrix $(\det{\Lambda})^2=1$, there are two classes of rotations with $\det{\Lambda}=\pm1$. 
In the first case, we have a subgroup $\rm SO(3)$ of the group ${\rm O(3)}$.

Consider the following statement: Any matrix $U\in{\rm SU(2)}$ can be represented in the form
$$
	U=U(\varphi,\theta,\psi)=U_z(\psi) U_y(\theta) U_z(\varphi)
$$
with matrices
$$
	U_z(\alpha)=\begin{pmatrix} \exp(i\alpha/2) & 0 \\ 0 & \exp(-i\alpha/2) \end{pmatrix}, \quad
	U_y(\alpha)=\begin{pmatrix} \cos(\alpha/2) & \sin(\alpha/2) \\ -\sin(\alpha/2) & \cos(\alpha/2) \end{pmatrix},
$$
where $\{\theta\in[0;\pi),\varphi\in[0;2\pi),\psi\in[-\pi;\pi)\}$ are the Euler angles.
To prove his statement, it is enough to show the relation between the ${\rm SU(2)}$ and ${\rm SO(3)}$ groups. 
The group ${\rm SU(2)}$ is locally isomorphic to the group ${\rm SO(3)}$, or equivalently, ${\rm SU(2)}$ is a double cover of ${\rm SO(3)}$.

In this way, we can consider the application of the Euler-angle parameterization for the tomography of spin states. 
For basis $|0\rangle$ and $|1\rangle$ we rewrite (\ref{def}) in the form
$$
	\mathcal{T}_{0}(\varphi,\theta,\psi)=\mathrm{Tr}[|0\rangle\langle0| U_z(\psi) U_y(\theta) U_z(\varphi)\rho U^\dagger_z(\varphi)U^\dagger_y(\theta)U^\dagger_z(\psi)].
$$
Using $U^\dagger_z(\psi) |0\rangle\langle0| U_z(\psi)=|0\rangle\langle0|$, we get the following result:
$$
	\mathcal{T}_{0}(\varphi,\theta,\psi)
	=\mathrm{Tr}[U^\dagger_z(\varphi)U^\dagger_y(\theta) |0\rangle\langle0| U_y(\theta) U_z(\varphi)\rho]=
	\mathrm{Tr}[|n(\varphi,\theta)\rangle\langle n(\varphi,\theta)|\rho]\equiv\mathcal{T}_{0}(\varphi,\theta)=\mathcal{T}_{0}(\mathbf{n}),
$$
where
$$
	|n(\varphi,\theta)\rangle=U^\dagger_z(\varphi)U^\dagger_y(\theta) |0\rangle=\cos(\theta/2)\exp(-i\varphi/2)|0\rangle+\sin(\theta/2)\exp(i\varphi/2)|1\rangle.
$$
and 
$$
	 \mathbf{n}=\mathbf{n}(\varphi,\theta)=R(\varphi,\theta)\mathbf{k}=\{\sin\theta\cos\varphi, \sin\theta\sin\varphi, \cos\varphi\},
$$
where $\mathbf{k}=\{0,0,1\}$ and the standard rotation matrix
\begin{equation}\label{SO3E}
	R(\varphi,\theta)=\begin{pmatrix}
	\cos\varphi\cos\theta & -\sin\varphi & \cos\varphi\sin\theta \\
	\sin\varphi\cos\theta & \cos\varphi & \sin\varphi\sin\theta \\
	-\sin\theta & 0 & \cos\theta
	\end{pmatrix}
\end{equation}

From the physical point of view, the spin tomogram $\mathcal{T}_m$ is the probability to observe the spin projection $m$ on the axis defined by the Euler angles $\varphi$ and $\theta$.

In view of the notation of [\ref{Manko2}, \ref{Manko3}, \ref{Filippov}], the relation for reconstructing the density operator reads
\begin{equation}\label{reconst}
	\hat\rho=\sum_{m=-j}^{j}{\int_{\mathbb{S}^2}{\frac{d\mathbf{n(\theta,\varphi)}}{4\pi}}\,\mathcal{T}^{j}(m,\mathbf{n(\theta,\varphi)})\hat{\mathcal{D}}^{j}(m,\mathbf{n(\theta,\varphi)})}, 
\end{equation}
where $m=-j,-j+1,\dots,j$ and
$$
	\mathcal{D}^{j}(m,\mathbf{n}(\theta,\varphi)))=(-1)^{m_2'}\sum_{j_3=0}^{2j}\sum_{m_3=-j_3}^{j_3}(2j_3+1)^2\times
$$
$$
	\times\sum_{m_1,m_1',m_2'=-j}^{j}(-1)^{m_1}D_{0m_3}^{j_3}(\theta,\varphi,0)
	{j \qquad j \qquad j_3 \choose m_1 \quad -m_1 \quad 0}{j \qquad j \qquad j_3 \choose m_1' \quad -m_2' \quad m_3}|jm_1'\rangle\langle jm_2'|,
$$
with$D_{mn}^{j}(\alpha,\beta,\gamma)$ being the Wigner function (generalized spherical function; for definition, see [\ref{Manko3}, \ref{Manko4}, \ref{Filippov}]).
This is inverse mapping to (\ref{map}).
It is worth noting that integration over the group is given by $\int{\mu_H(du)}\to\int_{\mathbb{S}^2}{d\mathbf{n}(\theta,\varphi)/4\pi}$ where ${\mu_H(du)}$ is the Haar measure. 

Note, that the Euler angles in direct way related to the Cayley-Klein parameters used for tomogram representation in previous section
\begin{equation}\label{CKE}
	\alpha=\exp(i(\psi+\varphi)/2)\cos(\theta/2), \quad \beta=\exp(i(\psi-\varphi)/2)\sin(\theta/2).
\end{equation}

\subsubsection{Example: qubit state}

As an illustrative example, we consider here a density operator $\hat{\rho}_0$ for 1/2 spin particle (qubit). 
And density matrix of this state can be expressed via the set of three Stokes parameters $\mathbf{S}=\{S_x,S_y,S_z\}$:
\begin{equation}\label{StokesP}
	\rho^0(\mathbf{S})=
	\frac{1}{2}(I+S_x\sigma_x+S_y\sigma_y+S_z\sigma_z)=
	\frac{1}{2}(I+\langle\mathbf{S},\sigma\rangle),
\end{equation}
where
\begin{equation}\label{Pauli}
	I=\begin{pmatrix} 1 & 0 \\ 0 & 1 \end{pmatrix},  \quad
	\sigma_x=\begin{pmatrix} 0 & 1 \\ 1 & 0 \end{pmatrix}, \quad
	\sigma_y=\begin{pmatrix} 0 & -i \\ i & 0 \end{pmatrix}, \quad
	\sigma_z=\begin{pmatrix} 1 & 0 \\ 0 & -1 \end{pmatrix}, \quad
\end{equation}
are identity matrix and Pauli matrices correspondingly, we use angular brackets for standard product. 
Stokes parameters satisfy to inequality
$$
	S_x^2+S_y^2+S_z^2\le1,
$$
and the equality holds for the pure state. 

For qubit state we obtain spin tomogram
$$ 
	\mathcal{T}_{0}(\mathbf{n},\mathbf{S})=\frac{1}{2}(1+S_x\sin\theta\cos\varphi+S_y\sin\theta\sin\varphi+S_z\cos\theta)=\frac{1}{2}(1+\langle\mathbf{S}, \mathbf{n}(\varphi,\theta)\rangle).
$$

\begin{figure}
\begin{minipage}[h]{0.35\linewidth}
\center{\resizebox{1.0\columnwidth}{!}{\includegraphics{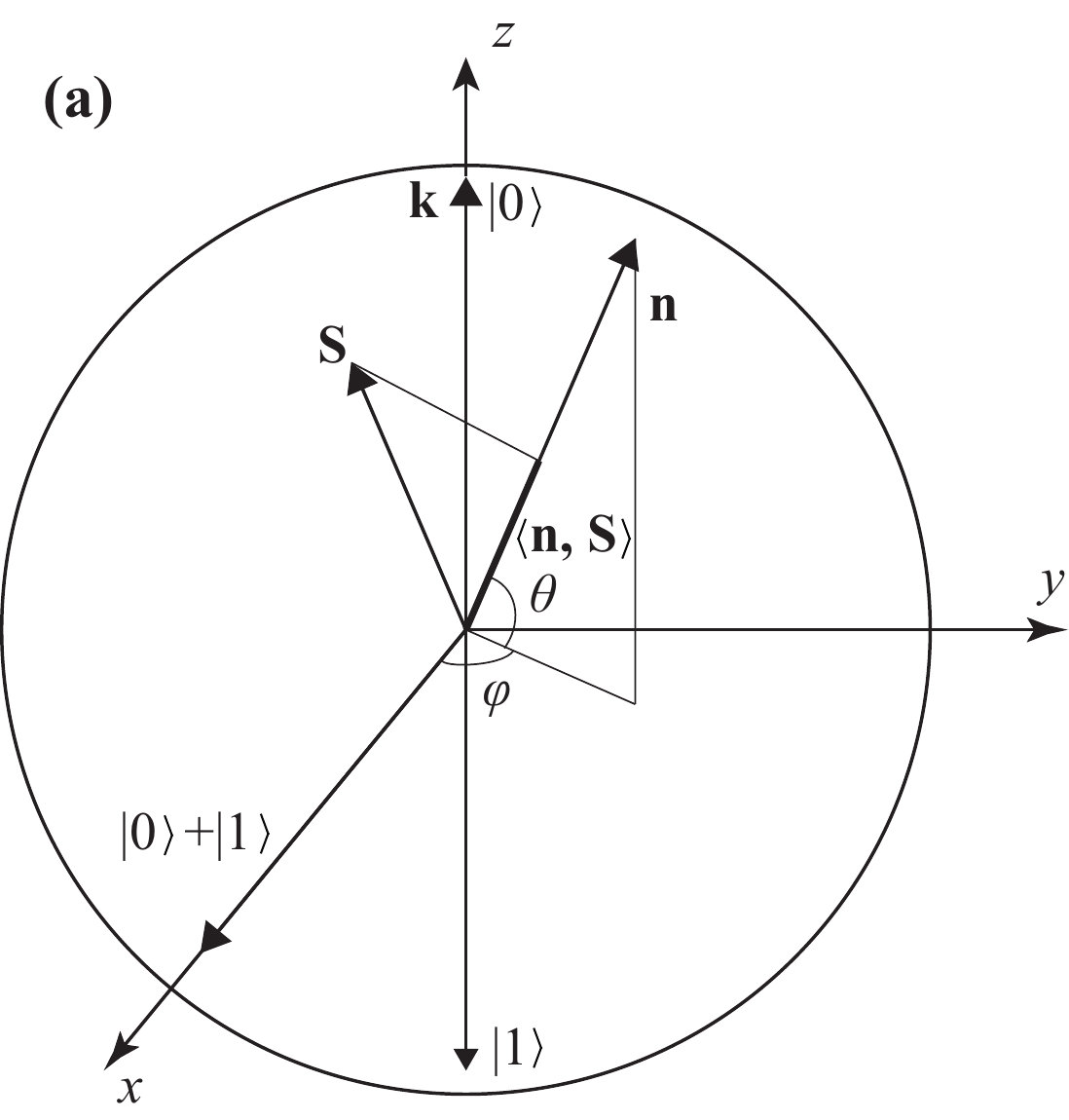}}}
\end{minipage}
\hfill
\begin{minipage}[h]{0.45\linewidth}
\center{\resizebox{1.0\columnwidth}{!}{\includegraphics{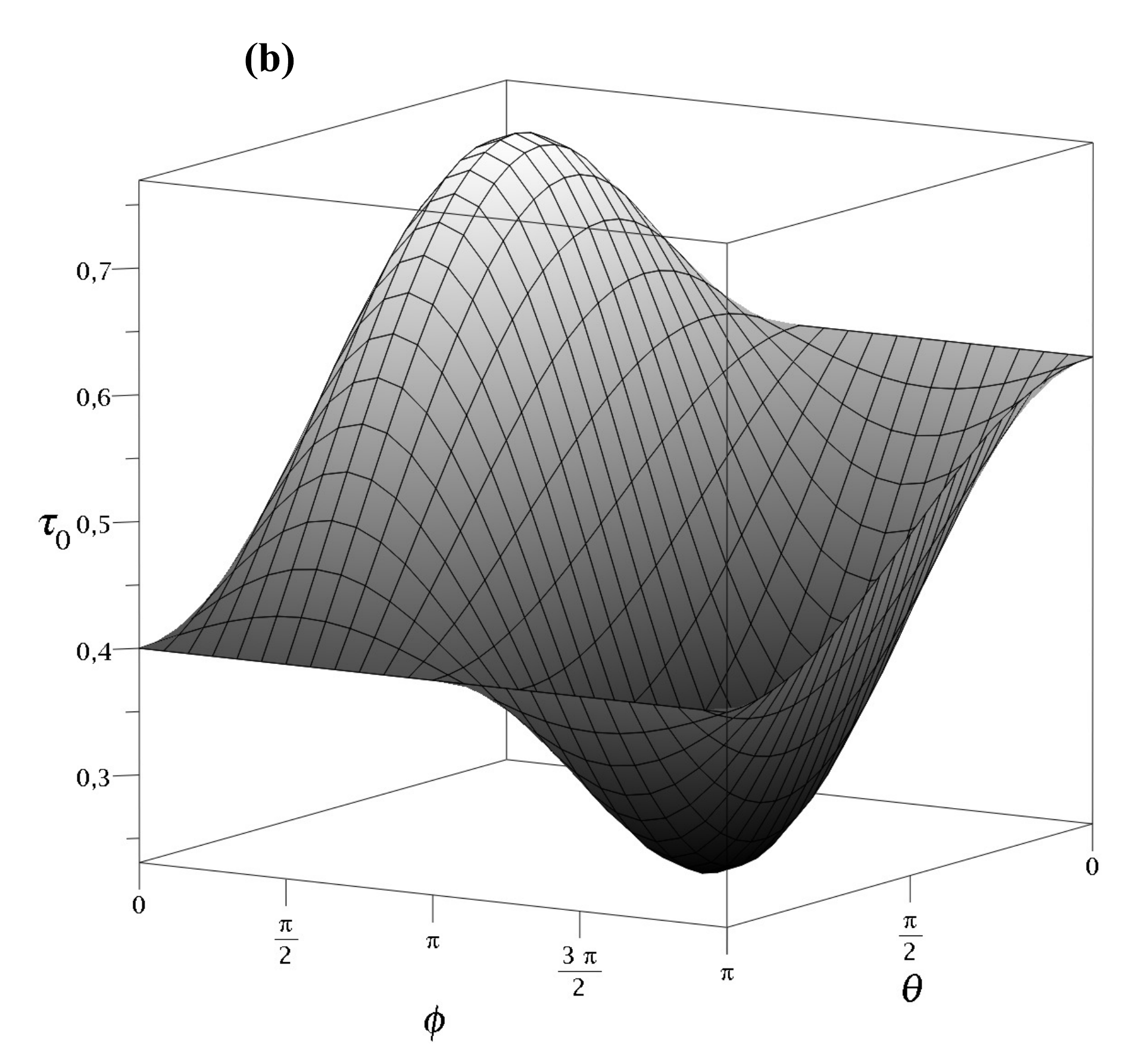}}}
\end{minipage}

\caption{
a) Geometrical interpretation of spin tomography scheme; 
b) Tomogram $\mathcal{T}_{0}$ for qubit state with $\mathbf{S}=\{0,0.5,0.2\}$ as function of Euler angles $\varphi$ and $\theta$.}
\label{fig:1}
\end{figure}

The geometrical interpretation of the spin tomography scheme is presented in Fig.\ref{fig:1}a. 
The tomogram for the qubit state with $\mathbf{S}=\{0,0.5,0.2\}$ is shown in  Fig.\ref{fig:1}b.

\section{Symplectic Spin Tomography}\label{Cayley}

Now we formulate our problem. 
There are two groups naturally connected with quantum tomography for continuous variables: $\rm{Sp}(2,\mathbb{R})$ and $\rm{SO}(2,\mathbb{R})$. 
In quantum tomography for discrete variables, the connection between ${\rm SU(2)}$ and ${\rm SO(3)}$ groups is used. 
We are interested in the construction of spin tomography based on the connection of the ${\rm SU(2)}$ group with some symplectic group.

First, we employ our mathematical intuition.

The simplest Lie group is the circle $\mathbb{S}\cong{\rm SO(2)}$, and extremely nice parameterization is given by the well-known Euler formula
$$
	e^{ix}=\cos{x}+i\sin{x}.
$$
This relation shows that the parameterization can be understood either in terms of the group of elements of norm $1$ in $\mathbb{C}$ [that is, the ${\rm U(1)}$ unitary group] or the imaginary subspace of $\mathbb{C}$.

Another compact Lie group is the sphere $\mathbb{S}^3\cong{\rm SU(2)}$. 
There exists a picture completely analogous to the previous argumentation, but with $\mathbb{C}$ replaced by the quaternions $\mathbb{H}$. 
We recall important issues related to quaternions. 
Let $\mathcal{E}$ be a linear space over a field $\mathbb{C}$ with the basis $\{e_0, e_1, e_2, e_3\}$, i.e., $\dim{\mathcal{E}}=4$. 
Let us introduce in $\mathcal{E}$ the following multiplication rules:
$$
	e_0^2=e_0, \quad e_0e_i=e_ie_0=e_i, \quad e_i^2=-e_0, \quad e_ie_j=\varepsilon_{ijk}e_k, \quad i=1,2,3,
$$
where $\varepsilon_{ijk}$ is the Levi-Civita symbol. 
The elements of the obtained ring (or algebra)  $\mathbb{H}$ are called quaternions. 
As follows from the definition, quaternion $\mathfrak{a}=(a_0,a_1,a_2,a_3)$ is
$$
	\mathfrak{a}=a_0e_0+a_1e_1+a_2e_2+a_3e_3+a_4e_4
$$
with $a_0=0$ corresponding to the vector and quaternion  $\mathfrak{a}$ with a $a_1=a_2=a_3=0$ corresponding to a scalar (we use gothic type for quaternions). 
As a real vector space, the quaternions are spanned by the four matrices
\begin{equation}\label{Quaternions}
	I=\begin{pmatrix} 1 & 0 \\ 0 & 1 \end{pmatrix},  \quad
	e_1=\begin{pmatrix} 0 & i \\ i & 0 \end{pmatrix}. \quad
	e_2=\begin{pmatrix} 0 & 1 \\ -1 & 0 \end{pmatrix}, \quad
	e_3=\begin{pmatrix} i & 0 \\ 0 & -i \end{pmatrix}, \quad
\end{equation}
In addition, unit quaternions form a group. 
This group is denoted ${\rm Sp(1)}$ since it is the first in the family of (compact) symplectic groups
$$
	{\rm Sp(1)}=\left\{\mathfrak{a}\in\mathbb{H}: |\mathfrak{a}|=\sqrt{a_0^2+a_1^2+a_2^2+a_3^2}=1\right\}.
$$
It is remarkable that the ${\rm Sp(1)}$ group is isomorphic to the group ${\rm SU(2)}$.

\subsection{Quaternion Representation}

In terms of quaternion $\mathfrak{a}$, we can get a new representation for spin tomograms
\begin{equation}\label{newdef}
	\mathcal{T}_m(\mathfrak{a})=\langle{m}|{U(\mathfrak{a})\rho U^{\dagger}(\mathfrak{a})}|m\rangle,
\end{equation}
where matrix $U\in{\rm SU(2)}$ is parameterized by unit quaternion $\mathfrak{a}$ in the form
$$
	U(\mathfrak{a})=\begin{pmatrix} a_0+ia_3 & a_2+ia_1 \\ a_2-ia_1 & a_0-ia_3 \end{pmatrix},
$$
as well as in the case of Euler-angles parameterization (\ref{SO3E}). 
This representation is similar to (\ref{defd}). 
Moreover, the reconstruction procedure is similar to (\ref{reconst}), as well.

Consider the relation for the density operator
\begin{equation}\label{newreconst}
	\hat\rho=\sum_{m=-j}^{j}{\int_{\mathbb{S}^3}{d\mathbf{n(\mathfrak{a})}}\,\mathcal{T}^{j}(m,\mathbf{n(\mathfrak{a})})\hat{\mathcal{D}}^{j}(m,\mathbf{n(\mathfrak{a})})}, 
\end{equation}
where we use the following transformations:
$$
	\int_{\mathbb{S}^2}{\frac{d\mathbf{n(\theta,\varphi)}}{4\pi}}\to{\int_{\mathbb{S}^3}d\mathbf{n(\mathfrak{a})}}, \quad
	\mathbf{n}(\varphi,\theta)=R(\varphi,\theta)\mathbf{k}\to\mathbf{n}(\mathfrak{a})=R(\mathfrak{a})\mathbf{k},
$$ 
with
$$
	R(\mathfrak{a})=
	\begin{pmatrix} 1-2a_2^2-2a_3^2 & 2a_1a_2-2a_0a_3 & 2a_1a_3+2a_0a_2 \\ 
	2a_1a_2+2a_0a_3& 1-2a_1^2-2a_3^2 & 2a_2a_3-2a_0a_1 \\ 
	2a_1a_3-2a_0a_2 & 2a_2a_3+2a_0a_1 & 1-2a_1^2-2a_2^2 
	\end{pmatrix}.
$$
We have the same relation for $\hat{\mathcal{D}}(m,\mathbf{n}(\mathfrak{a}))$ as in (\ref{reconst}) for spin tomography (see \cite{Filippov})
$$
	\mathcal{D}^{j}(m,\mathbf{n(\mathfrak{a})})=(-1)^{m_2'}\sum_{j_3=0}^{2j}\sum_{m_3=-j_3}^{j_3}(2j_3+1)^2\times
$$
$$
	\times\sum_{m_1,m_1',m_2'=-j}^{j}(-1)^{m_1}D_{0m_3}^{j_3}(\mathfrak{a})
	{j \qquad j \qquad j_3 \choose m_1 \quad -m_1 \quad 0}{j \qquad j \qquad j_3 \choose m_1' \quad -m_2' \quad m_3}|jm_1'\rangle\langle jm_2'|,
$$ 
where we transform from $D(\theta,\varphi)$ (see, [\ref{Manko3}, \ref{Manko4}, \ref{Filippov}]) to $D(\mathfrak{a})$ using standard relations for Euler angles
$$
	\varphi=\arctan_2\left(\frac{2(a_0a_1+a_2a_3)}{1-2(a_1^2+a_2^2}\right), \quad \theta=\arcsin{(2(q_0q_2-q_3q_1)}+\pi/2, \quad \psi=\arctan_2\left(\frac{2(a_0a_3+a_1a_2)}{1-2(a_2^2+a_3^2}\right),
$$
Here, the function $\arctan_2$ is used for  $\varphi$ with the addition of $2\pi$ to negative results. 
The function $\arctan_2(x,y)$ is defined as
$$
	\arctan_2\left(x,y\right)=\left\{\begin{array}{l}
	\arctan\left({\frac{y}{x}}\right),\quad x>0 \\ 
	\arctan\left({\frac{y}{x}}\right)+\pi, \quad y \ge 0,x < 0 \\ 
	\arctan\left({\frac{y}{x}}\right)-\pi, \quad y < 0,x < 0 \\ 
	+\frac{\pi}{2},\quad y > 0,x=0 \\ 
	-\frac{\pi}{2},\quad y < 0,x=0 \\ 
 {\rm{undef}},\quad y=x=0 \\ 
 \end{array} \right.
$$

In this case, the Haar measure reads
$$
	\mu(G)=\int_{B}{\frac{da_{0}da_{1}da_{2}da_{3}}{\det{G}}}=\int_{B}{\frac{d\mathfrak{a}}{(a_0^2+a_1^2+a_2^2+a_3^2)^2}}
$$
where $B$ is a Borel subset of $G$, and $d\mathfrak{a}=da_{0}da_{1}da_{2}da_{3}$ is the Lebesgue measure in $\mathbb{R}^4$. 
Note that in the case of a unit quaternion, $\det{G}=a_0^2+a_1^2+a_2^2+a_3^2=1$.

Directly from [\ref{Manko11}, \ref{Filippov}] we easily obtain the relation for purity of a state
\begin{equation}\label{purity}
	{\rm Tr}[\hat\rho^2]=
	(2j+1)\sum_{m}\int_{\mathbb{S}^3}{d\mathbf{n}(\mathfrak{a})\left(\sum_{m=-j}^{j}{\mathcal{T}^2(m,\mathbf{n})}-\sum_{m=-j}^{j-1}{\mathcal{T}(m,\mathbf{n}(\mathfrak{a}))\mathcal{T}(m+1,\mathbf{n}(\mathfrak{a}))}\right)},
\end{equation}
and the difference between the states $\rho_1$ and $\rho_2$ in the Hilbert--Smith metrics
\begin{equation}\label{diff}
	0\leq\max_{\mathbf{n}}\left(\frac{1}{2}\left[\mathcal{T}_1(m,\mathbf{n}(\mathfrak{a}))-\mathcal{T}_2(m,\mathbf{n}(\mathfrak{a}))\right]^2\right)^{1/2}\leq||\rho_1-\rho_2||_{HS},
\end{equation}
where $||x||_{HS}$ is the Hilbert--Smith metric \cite{Filippov}.

The quaternion parameters $a_0,a_1,a_2$ and $a_3$ are, in fact, obviously related to the Cayley--Klein parameters $\alpha_1=a_0$, $\alpha_2=a_3$, $\beta_1=a_2$, $\beta_2=a_4$.

\subsubsection{The Same Example: The Qubit State}

The tomogram of the qubit state reads
$$
	 \mathcal{T}_0(\mathfrak{a},\mathbf{S})=\frac{1}{2}(1+S_x(2a_1a_3+2a_0a_2)+S_y(2a_2a_3-2a_0a_1)+S_z(1-2a_1^2-2a_2^2))=
	 \frac{1}{2}(1+\langle \mathbf{S}, R(\mathfrak{a})\mathbf{k}\rangle).
$$
i.e., it is a function of three Stokes parameters $\{S_x,S_y,S_z\}$ and four quaternion parameters $\{a_0,a_1,a_2,a_3\}$. 
In spin tomography, the tomogram of the qubit state is a function of three Stokes parameters and two Euler angles $\{\theta,\varphi\}$. 
Therefore, as in the case of tomography of continuous variables, we have twice more parameters.

We conclude that we have constructed the representation for the spin tomogram for the qubit via the quaternion parameterization of the spin tomogram suggested above, and these relations are the main results of our work.

\subsubsection{Topological Argumentation}

We give topological arguments on the relation between ${\rm SU(2)}$, ${\rm SO(3)}$, and ${\rm Sp(1)}$ groups. 
The ${\rm SU(2)}$ group is topologically equivalent to the sphere $\mathbb{S}^3$ in $\mathbb{R}^4$, and the group ${\rm SO(3)}$ is topologically equivalent to the space $\mathbb{S}^3$.

The spin tomography representation is topologically based on the following connection. 
It is clear that elements of the ${\rm SU(2)}$  group are points of the 3-sphere $\mathbb{S}^3$ in  $\mathbb{R}^4$, where for any pair $\pm{U}\in{\rm SU(2)}$ we have diametral points on the sphere. 
In the case of homomorphism, these points become equivalent. 
Therefore, we have one of the models of $\mathbb{RP}^3$, i.e., the topological structure of the ${\rm SO(3)}$ group.

The quaternion representation is based on the simplest topological considerations. 
The argument is the following: The groups ${\rm SU(2)}$ and ${\rm Sp(1)}$ have explicitly the same topological structure of the 3-sphere $\mathbb{S}^3$ in the $\mathbb{R}^4$ space.

\subsubsection{Algebraic Argumentation: Pauli Matrices and Quaternions}

It should be noted, that algebra $A(\sigma)$ of the Pauli matrices (\ref{Pauli}) is isomorphic to quaternion algebra $\mathbb{H}$.
The isomorphism between them is given by
$$
	I=I, \quad i\sigma_k=e_k, \quad k=1, 2, 3.
$$
Therefore, the qubit state (\ref{StokesP}) is directly representable via quaternions
\begin{equation}\label{StokesP2}
\begin{split}
	\rho=
	\frac{1}{2}(I+S_x\sigma_x+S_y\sigma_y+S_z\sigma_z)
	\Leftrightarrow
	\rho=
	-\frac{i}{2}(e_0+S_xe_1+S_ye_2+S_ze_3)
\end{split}
\end{equation}
In this way, it does not matter how we transform the initial state to a tomographic representation, because all the tomographic representations are equivalent. 
In other words, we have a complete commutative diagram of transformations $A(\sigma)\leftrightarrow{\mathbb{H}}\leftrightarrow{\mathbf{n}(\theta,\varphi)}\leftrightarrow\mathbf{n}(\mathfrak{a})$.

\section{Conclusion}

In conclusion, we summarize the main results of our work.

We discussed the group-theoretical aspects of quantum tomography of continuous and discrete variables. 
We presented a new scheme for spin tomograms based on quaternion parameterization of the ${\rm SU(2)}$ group. 
This scheme is based on a new definition for tomogram (\ref{newdef}) and a new relation for reconstruction of the density operator (\ref{newreconst}). 
We considered the relation for the purity of a state (\ref{purity}) and difference between states in the Hilbert--Smith metrics (\ref{diff}). 
It should be noted that the relation is based on results adopted from [\ref{Manko11}, \ref{Filippov}].

Parameterization via the Euler angles does not come from a covering map of groups. 
At certain points, it has a problematic local behavior that is responsible for gimbal lock. 
An interesting and open question is how to construct the minimal scheme for symplectic spin tomograms in the spirit of the results of \cite{Filippov}

\section*{Acknowledgements}
The authors are thankful to Prof. V. I. Man’ko for useful comments and fruitful discussions, as well as to Dr. S. N. Filippov, Dr. A. I. Lvovsky, and Dr. A. I. Ovseevich for fruitful discussion. 
A.K.F. is an RQC Fellow, 
and E.O.K. was supported by the Russian Foundation for Basic Research under Projects Nos. 12-05-0000 and 12-05-98009 and the Council for Grants of the President of the Russian Federation (Grant No. SP-961.2013.5).


\begin{thebibliography}{9}
\bibitem{Shannon}\label{Shannon}
{C.~E.~Shannon},  Bell Syst. Tech. J., {\bf 27}, 379 (1948).

\bibitem{Shor}\label{Shor}
{P.~W.~Shor}, SIAM J. Computing, {\bf 26}, 1484 (1997).

\bibitem{Manko}\label{Manko}
S.~Mancini, V.I.~Man'ko, P.~Tombesi, Quantum Semiclass. Opt., {\bf 7}, 615 (1995).

\bibitem{Manko2}\label{Manko2}
S.~Mancini, V.I.~Man'ko, P.~Tombesi, Phys. Lett. A, {\bf 213}, 1 (1996).

\bibitem{Manko3}\label{Manko3}
V.~V.~Dodonov, V.~I.~Man’ko, Phys. Lett. A, {\bf 229} 335 (1997).

\bibitem{Manko4}\label{Manko4}
V.~I.~Manko, O.~V.~Man’ko, JETP, {\bf 85} 430 (1997).

\bibitem{Manko5}\label{Manko5}
V.~I.~Manko, O.~V.~Man’ko, S.~S.~Safonov, Theor. Math. Phys., {\bf 115}, 520 (1998).

\bibitem{Manko6}\label{Manko6}
V.~A.~Andreev, V.~I.~Man’ko, J. Opt. B: Quantum Semiclass. Opt., {\bf 2}, 122 (2000).

\bibitem{Manko7}\label{Manko7}
A.~S.~Arkhipov, et. al., Theor. Math. Phys. {\bf 142}, 311 (2005).

\bibitem{Manko8}\label{Manko8}
V.~N.~Chernega, V.~I.~Man'ko, J. Russ. Las. Res., {\bf 28}, 103 (2007).

\bibitem{Manko9}\label{Manko9}
A.~Ibort, et. al., Phys. Scr., {\bf 79}, 065013 (2009).

\bibitem{Manko10}\label{Manko10}
M.~A.~Man’ko, Theor. Math. Phys., {\bf 168}, 985 (2011).

\bibitem{Fedorov}\label{Fedorov}
A.~K.~Fedorov, S.~O.~Yurchenko, J. Phys.: Conf. Ser. {\bf 414}, 012040 (2013).

\bibitem{Fedorov2}\label{Fedorov2}
A.~K.~Fedorov, Phys. Lett. A, (2013).

\bibitem{Manko11}\label{Manko11}
S.~N.~Filippov, V.~I.~Man'ko. J. Russ. Las. Res., {\bf 34}, 14 (2013).

\bibitem{Beck}\label{Beck}
D.~T.~Smithey, et. al., Phys. Rev. Lett., {\bf 70}, 1244 (1993).

\bibitem{Lvovsky}\label{Lvovsky}
A.~I.~Lvovsky, M.~G.~Raymer, Rev. Mod. Phys., {\bf 81}, 299 (2009).

\bibitem{Lvovsky2}\label{Lvovsky2}
M.~Lobino, et. al., Science, {\bf 322}, 563 (2008).

\bibitem{Lvovsky3}\label{Lvovsky3}
A.~Anis, A.~I.~Lvovsky, New J. Phys., {\bf 14}, 105021 (2012).

\bibitem{Wigner}\label{Wigner}
E.~P. Wigner, Phys. Rev., {\bf 40}, 749 (1932).

\bibitem{Husimi}\label{Husimi}
E.~Husimi, Proc. Phys. Math. Soc. Jpn., {\bf 23}, 264 (1940).

\bibitem{Glauber}\label{Glauber}
R.~J.~Glauber, Phys. Rev. Lett., {\bf 10}, 84 (1963).

\bibitem{Sudarshan}\label{Sudarshan}
E.~C.~G. Sudarshan, Phys. Rev. Lett., {\bf 10}, 277 (1963).

\bibitem{Glauber2}\label{Glauber2}
K.~E.~Cahill, R.~J.~Glauber, Phys. Rev. A, {\bf 177}, 1882 (1969).

\bibitem{Hamilton}\label{Hamilton}
J.~J.~Hamilton, J. Math. Phys., {\bf 38}, 4914 (1997).

\bibitem{DAriano}\label{DAriano}
G.~M.~D'Ariano, L.~Maccone, M.~Paini, J. Opt. B: Quantum Semiclass. Opt., {\bf 5}, 77 (2003).

\bibitem{Filippov}\label{Filippov}
S.~N.~Filippov, ``Quantum states and dynamics of spin systems and electromagnetic field in the
tomographic-probability representation'', Ph.D. Thesis, Moscow Institute of Physics and Technology
[http://filippovsn.fizteh.ru/about/biography/Filippov-Abstract-Thesis.pdf (2012)].

\end{thebibliography}
\end{document}